\documentclass[10pt,twocolumn]{IEEEtran}
\usepackage{amsmath,amsthm,amssymb,epsfig,subfigure,cite}
\usepackage{algorithm,algpseudocode}

\newtheorem{theorem}{\hspace*{1pc}Theorem}

\begin{document}

\bibliographystyle{unsrt}

\setlength{\parindent}{1pc}

\title{Explicit Maximum Likelihood Loss Estimator in Multicast Tomography}
\author{Weiping~Zhu,\IEEEmembership{ member, IEEE,} \thanks{Weiping Zhu is with University of New South Wales, Australia}}
\date{}
\maketitle

\begin{abstract}
  For the tree topology, previous
studies show the maximum likelihood estimate (MLE) of a link/path
 takes a polynomial form with a degree that is one less than the
 number of descendants connected to the link/path. Since then, the main concern is focused on searching for methods to solve
 the high degree polynomial without using iterative approximation.
An explicit estimator based on the Law of Large Numbers has been
proposed
 to speed up the estimation.
 However, the estimate obtained from the estimator is not a MLE. When $n<\infty$, the estimate may be noticeable different from the MLE.
To overcome this, an explicit MLE estimator is presented in this paper
and a comparison between the MLE estimator and the explicit estimator
proposed previously is presented to unveil the insight of the MLE
estimator and point out the pitfall of the previous one.
\end{abstract}

\begin{IEEEkeywords}
Loss tomography, Tree topology, Transformation, Explicit Estimator.
\end{IEEEkeywords}

\section{Introduction}

Network tomography is proposed in \cite{YV96} to obtain network
characteristics without modifying network infrastructure, where the
author suggests the use of end-to-end measurement and statistical
inference together to estimate the characteristics instead of direct
measurement. The end-to-end measurement can be divided into two
classes: passive and active, depending on whether probe packets are
sent from sources to receivers. Without probing, the passive methods
depends on the data collected from log files to estimate network
characteristics. However, the data collected in the log files are
either unrelated or poorly related that makes inference hard if not
impossible. In contrast, the active approach attaches a number of
sources to some end nodes of a network that send probe packets to the
receivers attached to the other side of the network, where the paths
from the sources to the receivers cover the links of interest. Since
the probes are multicast to the receivers, the observations obtained
by the receivers are strongly correlated. Then, statistical inference
is applied on the data collected by the receivers to estimate the
network characteristics, such as link-level loss rates
\cite{Duffield2002}, delay distribution \cite{LY03}, \cite{TCN03},
\cite{PDHT02}, \cite{SH03}, \cite{LGN06}, and loss pattern
\cite{ADV07}. In this paper, our focus is on using active approach to
estimate the loss rate of a path/link.

Loss rate estimation is also called loss tomography in literature,
where the main focus is on searching for efficient maximum likelihood
estimators that can avoid the use of iterative procedures to
approximate the MLE. To achieve this, a deep analysis of the
likelihood function and a comprehensive study of the likelihood
equations obtained from the likelihood function are essential.
Unfortunately, there are only a few analytical results presented in
the literature, \cite{CDHT99} and \cite{Zhu06} are two of the few.
Both papers show that when the Bernoulli loss model is assumed for the
loss process of a link and independent identical distributed ({\it
i.i.d}) probing is used in end-to-end measurement, the maximum
likelihood equation of the pass/loss rate of a path/link takes a
polynomial form. The difference between them is that \cite{CDHT99} is
for the pass rate of a path connecting the source to an internal node,
\cite{Zhu06} is for the loss rate of a link connecting two nodes that
form a parent and child pair. We call them path-based estimator and
link-based estimator, respectively. Both estimators target the tree
topology, and their advantages and disadvantages are presented in
\cite{Zhu09}. Apart from agreeing on the polynomial form, both report
that the degree of the polynomial is one less than the number of
descendants connected to the path/link being estimated. Then, how to
solve a high degree polynomial becomes the critical issue since there
is no analytical solution for a polynomial that is 5 degree or greater
from Galois theory. Unfortunately, there has been little progress in
this regard until \cite{ZD09}, where a connection between observations
and the degree of the polynomial is established that provides the
theoretical foundation to reduce the degree of the polynomial obtained
from the likelihood equation. Prior to \cite{ZD09}, the authors of
\cite{DHPT06} introduced an explicit estimator built on the law of
large numbers. The estimator has been proved to be a consistent
estimator and has the same asymptotic variances as that of an MLE to
first order.  When $n < \infty$, the estimate obtained by the
estimator can be very different from the MLE. Considering the cost of
probing and dynamic nature of network traffic, we argue here that
despite the importance of the large sample theory in statistics, it is
unwise to use an estimator that is purely based on the law of large
numbers in practice because the accuracy of the estimator depends on a
large number of samples that can take a long time to collect and cost
a lot of resources.

The question then becomes whether there is an explicit maximum
likelihood estimator for multicast loss tomography. If so, what is
that and how different between the estimates obtained by the explicit
MLE estimator and the estimator presented in \cite{DHPT06} when $n <
\infty$. The two issues will be addressed in this paper. Firstly, we
present an explicit MLE estimator for the tree topology under the
Bernoulli model that has a similar computation complexity as the one
presented in \cite{DHPT06}. Secondly, a comparison between the two
estimators is presented that shows the newly proposed estimator is
better than the previous one when  $n < \infty$. The new estimator is
also better than the previous one in terms of the rate of convergence
when $n\rightarrow \infty$ since the MLE is asymptotic efficient and
the best asymptotically normal estimate.

By expanding both the statistical model used in the likelihood
equation and the observations obtained from receivers, we found that
the accuracy of an estimator is related to the consideration of the
correlated observations; while the efficiency of an estimator is
inversely related to the degree of the likelihood equation that is
proportional to the number of correlations. Then, how to keep the
accuracy without losing efficiency becomes  the key issue that has
been under investigation for some time. As a result, the connection
between
 the degree of the polynomial and the observations obtained by receivers is established
 that sets up the foundation to have an explicit maximum likelihood estimator. Meanwhile, the
exactly cause of the larger variance created by the explicit estimator
presented in \cite{DHPT06} is identified in the paper.

The rest of the paper is organized as follows. In Section 2, we
introduce the notations used in the paper. In addition to the
notation, the set of sufficient statistics used in this paper is
introduced in this section. We then present the explicit MLE in
Section 3 that unveils the connection between sufficient statistics
and the likelihood model used to describe the loss process of a
path/link. Section 4 is devoted to compare and analyze the explicit
MLE with the estimator presented in \cite{DHPT06}. The last section is
devoted to concluding remark.

\section{Problem Formulation}

\subsection{Notation}\label{treenotation}
In order to make correlated observations at the receivers, multicast
is used to send probes to receivers, where the multicast tree or
subtree used to connect the source to receivers is slightly different
from an ordinary one at its root, that has only a single child. Let
$T=(V, E)$ donate the multicast tree, where $V=\{v_0, v_1, ... v_m\}$
is a set of nodes representing routers and switches of a network;
$E=\{e_1,..., e_m\}$ is a set of directed links connecting node $f(i)$
to node $i$, where $f(i)$ is the parent of node $i$. To distinguish a
link from another, each link is assigned a unique number from 1 to m;
similarly, each node also has a unique number from 0 to m, where link
$i$ is used to connect node $f(i)$ to node $i$. The numbers are
assigned to the nodes from small to big along the tree structure from
top to bottom and left to right. The source is attached to node 0 to
send probes to the receivers attached to the leaf nodes of $T$. $R$ is
used to denote all receivers. Let $A=\{A_1,..., A_m\}$ be an m-element
vector, where $A_i, i \in \{1,\cdot\cdot,m\}$, is the pass rate of the
path connecting node 0 to node $i$. In addition, except leaf nodes
each node has a number of children, where $d_i$ denotes the children
of node $i$ and $|d_i|$ denotes the number of children of node $i$.
Note that a multicast subtree is different from an ordinary subtree,
where multicast subtree $i$, $T(i)$, is rooted at node $f(i)$ that has
link $i$ as its root link. The group of receivers attached to $T(i)$
is denoted by $R(i)$.  If $n$ probes are dispatched from the source,
each probe $i=1,...., n$ gives rise of an independent realization
$X^{(i)}$ of the probe process $X$, $X_k^i=1, k\in R$ if probe $i$
reaches receiver $k$; otherwise $X_k^i=0$. The observations of
$\Omega=(X^{(i)})^{i \in \{1,..,n\}}$ comprise the data for inference.

Given observation $X^{(j)}$, let

\begin{equation}
Y_i^{j}=\bigvee_{k \in R(i)} X_k^j, \mbox{\hspace{1cm}}  j \in \{1,
.., n\}.\label{projection0}\end{equation}

\noindent be the observation obtained by $R(i)$ for probe $j$. If
$Y_i^j=1$ probe $j$ reaches at least one receiver attached to $T(i)$,
that also implies the probe reaches node $i$. Then,
\[ n_i(1)=\sum_{j=1}^n Y_i^j, \] \noindent is the
number of probes confirmed from observations that reach node $i$.
$n_i(1), i \in V\setminus 0$ have been proved to be a set of minimal
sufficient statistics in \cite{ZD09}.

In addition to $n_i(1), i \in V\setminus 0$, we also introduced
another set of numbers for each node, $k$, where $n_{ij}(1) =
\sum_{u=1}^n (Y_i^u \wedge Y_j^u), i, j \in d_k $ is for the number of
probes confirmed from observations that reach at least one receiver of
$R(i)$ and one of $R(j)$ simultaneously; and $n_{ijk}(1)=\sum_{u=1}^n
(Y_i^u \wedge Y_j^u\wedge Y_k^u), i,j,k \in d_k $ is for the number of
probes confirmed from observations that reach simultaneously to at
least one receiver in each of $R(i)$, $R(j)$ and $R(k)$;
$\cdot\cdot\cdot$; and $n_{G}(1) = \sum_{u=1}^n (\bigwedge_{j \in G}
Y_j^u)$ is for the probes observed by at least one receiver in each
subtree rooted at node $k$.  We did not realize that this set of
numbers is a set of sufficient statistics until recently, we then name
them as the set of alternative sufficient statistics. The following
theorem confirms this:

\begin{theorem} \label{alternative set}
The alternative set of statistics defined above is a set of sufficient
statistics.
\end{theorem}
\begin{IEEEproof}
As stated, $n_i(1), i \in V\setminus 0$ has been proved to be a set of
minimal sufficient statistics. Then, if there is a function $\Gamma$
that can map the alternative set to $n_i(1), i \in V\setminus 0$, the
alternative set is a set of sufficient statistics. The function
$\Gamma$ is  as follows

\begin{eqnarray}
n_i(1)&=&\sum_{j \in d_i} n_j(1)-\sum_{\substack{ j<k\\
j, k \in d_i}} n_{jk}(1) \cdot\cdot + \nonumber \\
&&(-1)^{|d_i|-1} n_{d_i}(1), \mbox{     } i \in (V\setminus (0 \cup
R(i))) \nonumber \\
n_i(1)&=&\sum_{j=1}^n Y_i^j,   \mbox{    } i \in R
\end{eqnarray}
The function is a recursive function from bottom up along the tree
topology.
\end{IEEEproof}

\section{The Explicit Maximum Likelihood Estimator}
\subsection{Explicit Estimator}

Among the few studies providing analytical results, Multicast
Inference of Network Characters (MINC) is the most influence one that
covers almost all of the areas in network tomography, including
link-level loss, link delay and topology tomographies. In loss
tomography, it uses a Bernoulli model to describe the losses occurred
on a link. Using this model, the authors of \cite{CDHT99} derive an
MLE for the pass rate of a path connecting the source to an internal
node. The MLE is expressed in a set of polynomials, one for a path
\cite{CDHT99}, \cite{CDMT99}, \cite{CDMT99a}. Once knowing the pass
rates to two nodes that form a parent and child pair, the loss rate of
the link connecting the two node can be calculated by
$1-A_i/A_{f(i)}$. Considering the complexity of using numeric method
to solve higher degree polynomials $(
>5 )$, the authors of \cite{DHPT06}  propose an explicit estimator  on the basis of
the law of large numbers, where the authors define $Z_k^{(i)}= \min_{j
\in d_k} Y_j^{(i)}$ and $B_k=P(Z_k=1)$. The key of \cite{DHPT06} is
based on the following theorem.

\begin{theorem}

\begin{enumerate}
\item For $K \in V\setminus R$,
\begin{equation}
A_k=\Phi(B_k, \gamma):=\Big(\dfrac{\prod_{j \in
d_k}\gamma_j}{B_k}\Big)^{1/(|d_k|-1)} \label{explicit}
\end{equation}
\item Define $\breve{A}_k=\hat \gamma_k$ for $k \in R,$ and $\breve{A}_k=\Phi(\hat B_k,
\hat\gamma)$ otherwise. Then $\breve{A}_k$ is a consistent estimator
of $A_k$, and hence $\breve{A}_k=\breve{A_k}/\breve{A}_{f(k)}$ is a
consistent estimator of $A_k$.
\end{enumerate}
\end{theorem}

\noindent where $\hat B_k$, the empirical probability of $B_k$, is
equal to $n^{-1}\sum_{i=1}^n Z_l^{(i)}$. Note that
 the consistent property proved only ensures when $n\rightarrow \infty$, $\breve{A}_k$
is almost surely approach to $A_k$, the true pass rate of the path
from the source to node $k$. This property is the basic requirement
for an estimator. If an estimator cannot ensure consistency, it should
not be called an estimator. Then, the main concern with the explicit
estimator is its accuracy in comparison with the MLE when $n<\infty$
sine it only uses a part of all
 available
 information.

\subsection{Insight of MLE}

A minimum variance and unbiased estimator (MVUE) is normally regarded
as a good estimator in statistics. A maximum likelihood estimator is a
MVUE if it meets some simple regularity conditions. Unfortunately, the
estimator proposed in \cite{DHPT06} is not a MLE. Apart from that,
there are a number of concerns with the applicability of the estimator
in practice because the accuracy of the estimate requires
$n\rightarrow \infty$. Then, the scalability of the estimator must be
addressed, where the time and resources spent on measurement, time
spent on processing the data collected from measurement, and the
stationary period of network traffic must be considered in practice.

To remedy the scalability of the explicit estimator, we start to
search for an explicit maximum likelihood estimator and have a close
look at the maximum likelihood estimator proposed in
 \cite{CDHT99}, which is as follows:

\begin{eqnarray}
H(A_k, k) = 1-\dfrac{\gamma_k}{A_k} - \prod_{j \in
d_k}(1-\dfrac{\gamma_j}{A_k})=0 \label{treepoly}
\end{eqnarray}

 \noindent where $\gamma_i, i \in V\setminus 0$ is the pass rate of the multicast tree with its root link connecting node $0$ to node $i$.
 Rewriting (\ref{treepoly}) as
 \begin{eqnarray}
1-\dfrac{\gamma_k}{A_k}=\prod_{j \in d_k}(1-\dfrac{\gamma_j}{A_k}),
\label{treepoly1}
\end{eqnarray}
we found two interesting features of the polynomial; one is the
expandable feature, the other is merge-able feature. The former allows
us to expand both sides of (\ref{treepoly1}) to have a number of terms
on each side that are corresponded to each other. The latter, on the
other hand, allows us to merge a number of terms located on the right
hand side (RHS) and in the product into a single term as the one
located on left hand side (LHS) of (\ref{treepoly1}). The advantage of
the merge-able feature will be detailed in the next subsection. We now
put our attention on the expanding feature to unveil the internal
correlation embedded in (\ref{treepoly1}). Expanding the RHS of
(\ref{treepoly1}) and dividing all terms by $A_k$, we have

\begin{eqnarray}
\gamma_k&=&\sum_{j \in d_k}\gamma_j - \sum_{\substack{ j<k\\
j, k \in d_k}}\dfrac{\gamma_j\gamma_k}{A_k} \cdot\cdot \nonumber
\\ && + (-1)^{|d_k|-1}\dfrac{\prod_{j \in
d_k}\gamma_j}{A_k^{|d_k|-1}}. \label{mle}
\end{eqnarray}
Using the empirical probability $\hat \gamma_j = \dfrac{n_j(1)}{n}$ to
replace $\gamma_j$ in (\ref{mle}), we have a $|d_k|-1$ degree
polynomial of $A_k$. Solving the polynomial, the MLE of path $i$,
$\hat A_k$, is obtained. Note that $\hat\gamma_k$ can be replaced by
the alternative sufficient statistics, then the LHS of (\ref{mle})
becomes
\begin{multline}
\dfrac{1}{n}\big(\sum_{j \in d_k} n_j(1)-\sum_{\substack{ j<k\\
j, k \in d_k}} n_{jk}(1) \cdot\cdot + (-1)^{|d_k|-1} n_{d_k}(1) \big )
 \label{numberside}
\end{multline}
Comparing the RHS of (\ref{mle}) with (\ref{numberside}), one is able
to find the correspondences between the terms. Each term of
(\ref{mle}) represents a type of correlation in the model
among/between the members of the term, while each term of
(\ref{numberside}) is the statistics or evidence obtained from an
experiment for the corresponding term of (\ref{mle}). Except the first
term of (\ref{mle}) and the first term of (\ref{numberside}) that are
exactly equal to each other, other pairs between the two can be
different from each other if $n<\infty$. Taking the first terms of
(\ref{mle}) and (\ref{numberside}) out, we have
\begin{eqnarray}
&&\sum_{\substack{ j<k\\
j, k \in d_k}}\dfrac{\gamma_j\gamma_k}{A_k} \cdot\cdot +
(-1)^{|d_k|-1}\dfrac{\prod_{j \in d_k}\gamma_j}{A_k^{|d_k|-1}} \nonumber \\ &=&\dfrac{1}{n}(\sum_{\substack{ j<k\\
j, k \in d_k}} n_{jk}(1) \cdot\cdot + (-1)^{|d_k|-1} n_{d_k}(1))
\label{mle2num}
\end{eqnarray}
Statistical inference aims to estimate $\hat A_k$  from
(\ref{mle2num}), i.e. matching the model presented on the LHS to the
statistics presented on the RHS. This equation also shows that in
order to have the MLE of a path, one must consider all available
information embedded in the observations of $R(i)$, in particular the
correlations between the descendants. Without correlations and/or the
corresponding statistics, inference is impossible. This corresponds to
the data consistent problem raised in \cite{CDHT99} and \cite{Zhu09}.
If we only consider matching a part of (\ref{mle}) to the
corresponding part of (\ref{numberside}) when $n<\infty$, the estimate
obtained would not be the MLE unless the ignored correlations are
negligible. Then, the explicit estimator proposed in \cite{DHPT06} is
not an MLE since it only pairs the last term of (\ref{mle}) to the
last term of (\ref{numberside}).

\subsection{The Explicit Maximum Likelihood Estimator}

As stated, the degree of the polynomial is proportional to the number
of descendants connected to the path being estimated and the
estimation relies on the observed correlations between the descendants
to estimate the unknown characteristic.  Under the i.i.d. assumption,
the likelihood function takes a product form as (\ref{treepoly1}).
Unfortunately, the previously stated merge-able feature has not been
given enough attention although (\ref{treepoly1}) clearly expresses
the loss rate of subtree $k$ is equal to the product of the loss rates
of those sub-multicast trees rooted at node $k$.  In probability,
(\ref{treepoly1}) states such a fact that the loss rate of subtree $k$
depends on a number of independent events, one for a sub-multicast
tree rooted at node $k$. More importantly, this implies that those
independent events can be merged into a single event, i.e. the LHS of
(\ref{treepoly1}). With this in mind, whether the degree of
(\ref{treepoly}) can be reduced depends on whether we are able to
obtain the empirical pass rate of the tree that has a path from the
source to node $k$ plus {\it some} of the multicast subtrees rooted at
node $k$. Let $\gamma_{k_g}$ denote the pass rate, where $k$ is for
the end node of the path being estimated, $g$ denotes the group of
subtrees being merged. Based on the alternative sufficient statistics
of $d_k$,   $\hat\gamma_{k_g}$, the empirical probability of
$\gamma_{k_g}$, can be computed
 \cite{ZD09}.
Then, the degree of (\ref{treepoly1}) can be reduced to 1 that can be
solved easily. Further, we have the following theorem to calculate the
pass rate of a path explicitly for the tree topology.

\begin{theorem} \label{explicit MLE}
For the tree topology that uses the Bernoulli model to describe the
loss process of a link, there is an explicit MLE estimator to estimate
the pass rate of the path connecting the source to node $k, k \in
V\setminus (0\cup R)$, which is as follows:
\begin{equation}
\hat
A_k=\dfrac{\hat\gamma_{k1}\hat\gamma_{k2}}{\hat\gamma_{k1}+\hat\gamma_{k2}-\hat\gamma_k}
\end{equation}
where $\hat\gamma_k=\dfrac{n_k(1)}{n}$,
$\hat\gamma_{k1}=\dfrac{n_{k1}(1)}{n}$, and
$\hat\gamma_{k2}=\dfrac{n_{k2}(1)}{n}$. $n_{k1}(1)$ and $n_{k2}(1)$
are the number of probes confirmed from observations reaching at least
one receiver attached to the merged subtree 1 and 2, respectively.
\end{theorem}
\begin{IEEEproof}
Since node $k$ is not a leaf node, the subtrees rooted at the node can
be divided into two exclusive groups: $d_{k1}$ and $d_{k2}$ where
$d_{k1} \cup d_{k2}=d_k$ and $d_{k1} \cap d_{k2}=\emptyset$. The
statistics of the merged subtrees can be computed by using $d_{k1}$ or
$d_{k2}$ to replace $d_k$ in (\ref{numberside}). Then, we have
\begin{eqnarray}
1-\dfrac{\hat\gamma_k}{A_k}&=&\prod_{j \in
d_k}(1-\dfrac{\hat\gamma_j}{A_k}) \nonumber
\\
&=& \prod_{j \in d_{k1}}(1-\dfrac{\hat\gamma_j}{A_k} )\prod_{k \in d_{k2}}(1-\dfrac{\hat\gamma_k}{A_k}) \nonumber \\
&=& (1-\dfrac{\hat\gamma_{k1}}{A_k})(1-\dfrac{\hat\gamma_{k2}}{A_k})
\label{betalinktogamma}
\end{eqnarray} Solving (\ref{betalinktogamma}), we have

\[
A_k=\dfrac{\hat\gamma_{k1}\hat\gamma_{k2}}{\hat\gamma_{k1}+\hat\gamma_{k2}-\hat\gamma_k}
\]

\end{IEEEproof}

The theorem shows that using (\ref{numberside}) to merge the
alternative statistics of multiple multicast subtrees rooted at the
same node would not affect the estimation of the pass rate of the path
that ends at the node.

\section{Comparison of Estimators}

In this section, we tackle the second task set at the beginning of the
paper, i.e. compare the explicit MLE estimator against the explicit
one presented in \cite{DHPT06} for $n <\infty$. Two scenarios: 2
descendants and 3 descendants are connected to the path being
estimated, are considered to illustrate the estimates received from
the explicit estimator proposed in \cite{DHPT06} drifts away from the
MLE as the number of descendants increases.
 To make the variance approximate to the first order of the MLE, the explicit estimator needs to send more probes to receivers.
  We will use $H(A_i,i)$ as a
reference in the following comparison to measure the accuracy between
an estimator and its MLE counterpart.
\subsection{Binary Tree}

\begin{figure}
\centerline{\psfig{figure=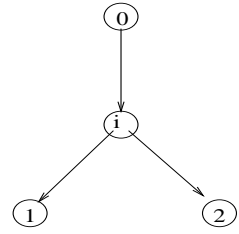,height=3.0cm,width=3cm}}
\caption{A Binary Tree} \label{binary}
\end{figure}

For a tree with binary descendants as  Figure \ref{binary}, the pass
rate of link/path i estimated by $H(A_i,i)$ is equal to
\begin{eqnarray}
\hat A_i &=&
\dfrac{\hat\gamma_2\hat\gamma_1}{\hat\gamma_2+\hat\gamma_1-\hat\gamma_i}
\end{eqnarray}
\noindent Using the explicit estimator of \cite{DHPT06}, we have

\begin{eqnarray}
\breve{A}_i&=&\dfrac{\hat\gamma_2\hat\gamma_1}{\dfrac{n_{12}(1)}{n}}
\nonumber
\\
&=&\dfrac{\hat\gamma_2\hat\gamma_1}{\dfrac{n_2(1)+n_1(1)-n_i(1)}{n}}
\nonumber \\
&=&
\dfrac{\hat\gamma_2\hat\gamma_1}{\hat\gamma_2+\hat\gamma_1-\hat\gamma_i}
\label{2node}
\end{eqnarray}

\noindent It is the same as the MLE. This is because  there is only
one type of correlation between the model and the observations, which
are considered by the estimator. Thus, $\breve{A}_i=\hat A_i$.

Based on Theorem \ref{explicit MLE}, the estimate obtained by the
estimator proposed in this paper is also the same as above, which can
be written as:
\begin{eqnarray}
\hat A_i &=& \dfrac{n_1(1)n_2(1)}{n\cdot (n_1(1)+n_2(1)-n_i(1))}
\nonumber \\
&=&
\dfrac{\dfrac{n_1(1)n_2(1)}{{n^2}}}{\dfrac{n_1(1)+n_2(1)-n_i(1)}{n}}.
\end{eqnarray}

Thus, for the binary tree, the three estimators produce the same
result.
\subsection{Tertiary Tree}

Let $i$ have three descendants 1, 2, and 3. Based on $H(A_i, i)$, we
have
\begin{eqnarray}
&& \hat A_i^2(\hat\gamma_1+\hat\gamma_2+\hat\gamma_3 -
\hat\gamma_i)-\nonumber
\\ && \hat A_i(\hat\gamma_1\hat\gamma_2+\hat\gamma_1\hat\gamma_3+\hat\gamma_2\hat\gamma_3)
+\hat\gamma_1\hat\gamma_2\hat\gamma_3=0. \nonumber \\\label{3nodemle}
\end{eqnarray}

\noindent Solving the quadratic function, we have the MLE of $\hat
A_i$

Based on (\ref{mle2num}), the model and observations are connected by

\begin{multline} \sum_{\substack{ j<k\\ j, k \in \{1,2,3\}}}
\dfrac{n_{jk}(1)}{n} -
 \dfrac{n_{123} (1)}{n}= \\ \sum_{\substack{ j<k
\\ j, k \in \{1,2,3\}}}\dfrac{\gamma_j\gamma_k}{\hat A_i}
 - \dfrac{\gamma_1\gamma_2\gamma_3}{\hat A_i^2} \label{3children}
 \end{multline}

It is easy to prove (\ref{3children}) equals (\ref{3nodemle}). Using
theorem \ref{explicit MLE}, we have the MLE directly,

\begin{eqnarray}
\hat A_i = \dfrac{(n_1(1)+n_2(1)-n_{12}(1))\cdot
n_3(1)}{n\cdot(n_{13}(1)+n_{23}(1)-n_{123}(1))}. \label{3mle}
\end{eqnarray}

\noindent Intuitively, one can notice that (\ref{3mle}) considers all
correlations between the 3 descendants. To prove this equals
(\ref{3nodemle}), we write the RHS of the above as
\begin{equation}
\dfrac{\dfrac{(n_1(1)+n_2(1)-n_{12}(1))}{n}\cdot \dfrac{
n_3(1)}{n}}{\dfrac{(n_{13}(1)+n_{23}(1)-n_{123}(1))}{n}}.
\label{3equality}
\end{equation}
The denominator of the above is obtained from

\begin{eqnarray}
&&\dfrac{n_1(1)+n_2(1)-n_{12}(1)}{n}+\dfrac{n_3(1)}{n} - \gamma_i.
\label{denominator}
\end{eqnarray}
According to (\ref{mle}) and (\ref{numberside}),
\[
\gamma_1+\gamma_2-\dfrac{\gamma_1\gamma_2}{\hat
A_i}=\dfrac{n_1(1)+n_2(1)-n_{12}(1)}{n}
\]
Using the above in (\ref{denominator}), the denominator turns into
\begin{eqnarray}
(\gamma_1+\gamma_2+\gamma_3-\gamma_i)-\dfrac{\gamma_1\gamma_2}{\hat
A_i}. \label{deno}
\end{eqnarray}
Similarly, the nominator of (\ref{3equality}) is equal to
\begin{eqnarray}
(\gamma_1+\gamma_2-\dfrac{\gamma_1\gamma_2}{\hat A_i})\cdot\gamma_3 =
\gamma_1\gamma_3+\gamma_2\gamma_3-\dfrac{n_{12}(1)}{n}\gamma_3.
\nonumber \\ \label{nomi}
\end{eqnarray}
Using (\ref{deno}) and (\ref{nomi}) to replace the denominator and
nominator of (\ref{3mle}), we have
\begin{eqnarray}
&&\hat A_i\cdot(\gamma_1+\gamma_2+\gamma_3-\gamma_i)-\hat
A_i\cdot\dfrac{\gamma_1\gamma_2}{\hat A_i}
\nonumber \\
&=&\gamma_1\gamma_3+\gamma_2\gamma_3-\dfrac{\gamma_1\gamma_2}{\hat
A_i}\gamma_3
\end{eqnarray}
Moving every term to the LHS and multiplying $\hat A_i$, we have
(\ref{3nodemle}).

 Because of the symmetric nature of the 3 descendants, we can
also merge descendants 2 and 3 first or merge descendants 1 and 3
first, that lead to

\[
\hat A_i = \dfrac{(n_2(1)+n_3(1)-n_{23}(1))\cdot
n_1(1)}{n\cdot(n_{13}(1)+n_{12}(1)-n_{123}(1))}
\]

\noindent and
\[
\hat A_i = \dfrac{(n_1(1)+n_3(1)-n_{13}(1))\cdot
n_2(1)}{n\cdot(n_{12}(1)+n_{23}(1)-n_{123}(1))}
\]
respectively.

In contrast, the explicit estimator presented in \cite{DHPT06} has its
estimate
\begin{equation}
\breve{A_i}=\Big(\frac{\gamma_1\gamma_2\gamma_3}{\dfrac{n_{123}(1)}{n}}\Big)^{\frac{1}{2}}.\label{expli}
\end{equation}

\noindent Comparing (\ref{expli}) with (\ref{3mle}), a direct
impression is that (\ref{expli}) fails to match the paired
correlations, i.e. descendants 1 and 2, descendants 1 and 3, and
descendants 2 and 3. If we assume (\ref{expli}) is a solution of a
quadratic equation, the equation should be as follows:
\begin{multline}
 \dfrac{n_{123}(1)}{n}\breve{A_i}^2 -
2\cdot\big(\dfrac{n_{123}(1)\gamma_1\gamma_2\gamma_3}{n}\big)^{\frac{1}{2}}\breve{A_i}+\gamma_1\gamma_2\gamma_3=0
\label{3expli}
\end{multline}
that has a double root. Given the double root assumption,
(\ref{3expli}) is certainly not the polynomial that leads to MLE since
it is contradict to the Lemma 1 introduced in \cite{CDHT99} and
\cite{ZD09} that states there is one and only one root in $(0,1)$ for
the maximum likelihood equation. Then, the estimator proposed in \cite
{DHPT06} can be regarded as an estimator based on the method of
moments that will be discussed in the next subsection.

\subsection{Analysis}

The fundamental principle of maximum likelihood estimate is unveiled
clearly by (\ref{mle2num}), where the LHS of (\ref{mle2num}) is the
statistical distribution of the loss process (called model previously)
and the RHS is the statistics obtained from observations. There is one
to one correspondence between the terms cross the equal sign. The
maximum likelihood estimator aims to solve the equation to find the
$A_k$ that fits to the statistic model; while the explicit estimator
proposed previously attempts to use the last terms of both sides. When
$n < \infty$, the estimate can be different from MLE. In fact, if only
considering asymptotic accuracy, each of the corresponding terms can
be connected and considered an explicit estimator as the one presented
in \cite{DHPT06}. All of the explicit estimators can also be proved to
be consistent as their predecessor. Then, the theorem 1 can be
extended as
\begin{theorem} \label{set of estimators}
Each of the corresponding pairs of (\ref{mle2num}) forms an explicit
estimator that is consistent as the one proposed in \cite{DHPT06},
that has the form of
\begin{eqnarray}
\Phi_i(n_{w}(1), w, \gamma)=A_i= \dfrac{1}{C_{|w|}^{|d_i|}}\sum_{ j
\in w}\Big (\dfrac{\prod_{k\in
j}\gamma_k}{\frac{n_{w}(1)}{n}}\Big)^{\frac{1}{|w|-1}}
\end{eqnarray}
where $w$ corresponds to one of the pairs cross the equal sign of
(\ref{mle2num}) and $|w| < |d_i|$ denotes the number of members in the
term, and $n_{w}(1)$ corresponds to the statistics denoting the number
of probes reaching the receivers attached to the descendants of $w$.
\end{theorem}
\begin{IEEEproof}
It can be proved as the proof of Theorem 2 in \cite{DHPT06}.
\end{IEEEproof} Clearly, when
$n<\infty$, the estimates obtained by the explicit estimators are not
the MLE. Note that each term on the RHS of (\ref{mle2num}) is not a
sufficient statistic by itself but only a part of the sufficient
statistics defined in Theorem \ref{alternative set}. Combining a
number of the estimators defined above can improve the accuracy of the
estimate. When all terms are combined, we have the MLE.

Despite having an explicit maximum likelihood estimator, we still
carry out the following analysis to find out why the partial matching
could not be an MLE. Firstly, we compare (\ref{mle2num}) with the
explicit one proposed in \cite{DHPT06} on the basis of polynomial. The
former is a $|d_k|-1$ degree of polynomial that has a unique solution
in $(0,1)$, while the latter can be considered a polynomial that has a
multiple root in $(0,1)$. In other wards, the explicit one considers
the sum of the first $|d_k|-2$ terms on both sides of the equal sign
are equal to each other, so does the last terms. Instead of using the
sum of the first $|d_k|-2$ terms to estimate $A_k$, the explicit
estimator uses the last one to avoid solving a $|d_k|-2$ degree
polynomial.  Using this approach to estimate $A_k$, an error is
inevitable and the amount of the error depends on the number of
descendants; and the more the better. This is because if a node has
more descendants, we are able to obtain more information about the
path connecting the source to the node from observations.

 The explicit estimator proposed in \cite{DHPT06} can also be viewed as a method of moments. If
so, its estimate is normally superseded by the maximum likelihood,
because maximum likelihood estimators have higher probability of being
close to the quantities to be estimated. Also, the estimates obtained
by the method of moments are not necessarily based on sufficient
statistics, i.e., they sometimes fail to take into account all or a
large part of the relevant information in the sample. Therefore, the
accuracy of such an estimator depends on large sample. As stated,
$n_{d_k}(1)$ itself is not a sufficient statistic for $A_k$, using
$n_{d_k}(1)$ alone to estimate $A_k$ fails to consider all other
correlations between descendants. Then, as stated, error is
inevitable. Let $n$ be the sample size and $\delta$ be the error rate,
their relation is expressed as

\[
\dfrac{\delta}{\sqrt{n}} \]
 Based on the formula, the explicit
estimator relies on sending infinite number of probes to reduce the
error. Even though, the effect of ignoring other correlations remains,
that makes the variance of the explicit estimator can only approximate
to the first order of that of the maximum likelihood estimator.

\subsection{Computational Complexity}

The estimator proposed in this paper is the maximum likelihood one
with a similar computation complexity as that of the explicit
estimator presented in \cite{DHPT06}. To determine $\hat A_k$ or
$\breve{A_k}$, both need to calculate the empirical probabilities
$\hat \gamma_i, i \in V\setminus 0$. The two estimators require to
compute $Y_j^i$ and $n_j(1)$, there are total $O(n\cdot(|V|-1))$
operations. In addition, the estimator proposed here needs to merge
descendants into two for those nodes that have more than 2
descendants. That requires to compute $n_{jk}(1), ..., n_{d_k/2}(1)$
for node $k$, and there are $2\times
(2^{\frac{|d_k|}{2}}-\dfrac{|d_k|}{2}-1)$ operations. On the other
hand, the previous explicit estimator needs to calculate $\hat B_k$
that requires the computation of $Z_k^i$ for each node that takes a
smaller amount of operations than the MLE does. On the other hand, the
explicit one needs to perform n-th root operation for each node that
has more than two children, while the MLE one only needs a simple
arithmetic operation to estimate $A_k$. Therefore, in terms of
operations, the two are similar to each other.

\section{Conclusion}

In this paper, an explicit MLE is proposed that is built on the unique
features of the likelihood equation and the set of alternative
sufficient statistics introduced in this paper. The two features of
the likelihood equation, i.e. expandable and merge-able, can be
considered micro and a macro views of the
 likelihood equation. The rise of the macro view makes merging
 possible; and the rise of the micro view unveils the fundamental
  of the explicit estimator proposed previously and the internal
  correlations in the model and observations. Based on the macro view,
  a closed form MLE estimator is proposed and presented in this paper,
  which is of the simplest one that
 has even been presented in literature. Applying the micro view on
 (\ref{treepoly1}),
we establish the correlations between descendants, and we also
establish the correspondence between the statistical model and the
statistics obtained from the leaf nodes of the descendants. This
correspondence further unveils the connection between the observations
and the degree of the likelihood polynomial. As a result, the explicit
MLE is proposed for the tree topology. In addition to the explicit
estimator, we in this paper compare the estimator proposed in this
paper with the explicit estimator proposed previously, which shows
that when $n <\infty$, the MLE one is substantially better than the
explicit one in terms of accuracy.

\bibliography{zhu}

\end{document}